\documentclass[preprint,superscriptaddress,aps,longbilbiography]{revtex4-2}
\usepackage{xcolor}
\usepackage{graphicx}% Include figure files
\usepackage{bm}% bold math
\usepackage{amsmath}
\usepackage{amssymb}
\usepackage{color}
\begin{document}

\title{Revisiting altermagnetism in RuO\textsubscript{2}: a study of laser-pulse induced charge dynamics by time-domain terahertz spectroscopy}

\author{David T. Plouff}\email{dplouff@udel.edu}
\affiliation{Department of Physics and Astronomy, University of Delaware, Newark, DE 19716, USA}
\author{Laura Scheuer}%\email{scheuer@udel.edu}
\affiliation{Department of Physics and Astronomy, University of Delaware, Newark, DE 19716, USA}
\author{Shreya Shrestha}%\email{sthashre@udel.edu}
\affiliation{Department of Physics and Astronomy, University of Delaware, Newark, DE 19716, USA}
\author{Weipeng Wu}%\email{wpwu@udel.edu}
\affiliation{Department of Physics and Astronomy, University of Delaware, Newark, DE 19716, USA}
\author{Nawsher J. Parvez}%\email{parvez@udel.edu}
\affiliation{Department of Physics and Astronomy, University of Delaware, Newark, DE 19716, USA}
\author{Subhash Bhatt}%\email{bhattsb@udel.edu}
\affiliation{Department of Physics and Astronomy, University of Delaware, Newark, DE 19716, USA}
\author{Xinhao Wang}%\email{xinhaow@udel.edu}
\affiliation{Department of Physics and Astronomy, University of Delaware, Newark, DE 19716, USA}
\author{Lars Gundlach}\email{larsg@udel.edu}
\affiliation{Department of Physics and Astronomy, University of Delaware, Newark, DE 19716, USA}
\affiliation{Department of Chemistry and Biochemistry, University of Delaware, Newark, DE 19716, USA}
\author{M. Benjamin Jungfleisch}\email{mbj@udel.edu}
\affiliation{Department of Physics and Astronomy, University of Delaware, Newark, DE 19716, USA}
\author{John Q. Xiao}\email{jqx@udel.edu}
\affiliation{Department of Physics and Astronomy, University of Delaware, Newark, DE 19716, USA}

\date{\today}

\begin{abstract}
\textbf{Altermagnets are a recently discovered class of magnetic material with great potential for applications in the field of spintronics, owing to their non-relativistic spin-splitting and simultaneous antiferromagnetic order. One of the most studied candidates for altermagnetic materials is rutile structured RuO\textsubscript{2}. However, it has recently come under significant scrutiny as evidence emerged for its lack of any magnetic order. In this work, we study bilayers of epitaxial RuO\textsubscript{2} and ferromagnetic permalloy (Fe\textsubscript{19}Ni\textsubscript{81}) by time-domain terahertz spectroscopy, probing for three possible mechanisms of laser-induced charge dynamics: the inverse spin Hall effect (ISHE), electrical anisotropic conductivity (EAC), and inverse altermagnetic spin-splitting effect (IASSE). We examine films of four common RuO\textsubscript{2} layer orientations: (001), (100), (110), and (101). If RuO\textsubscript{2} is altermagnetic, then the (100) and (101) oriented samples are expected to produce anisotropic emission from the IASSE, however, our results do not indicate the presence of IASSE for either as-deposited or field annealed samples. The THz emission from all samples is instead consistent with charge dynamics induced by only the relativistic ISHE and the non-relativistic and non-magnetic EAC, casting further doubt on the existence of altermagnetism in RuO\textsubscript{2}. In addition, we find that in the (101) oriented RuO\textsubscript{2} sample, the combination of ISHE and EAC emission mechanisms produces THz emission which is tunable between linear and elliptical polarization by modulation of the external magnetic field.}
\end{abstract}

\maketitle

\newpage
\section{\label{sec:level1}Introduction\protect}

Altermagnetism is a newly identified magnetic phase characterized by collinear antiferromagnetism coexisting simultaneously with anisotropic spin-splitting in the band structure, owing to the two spin-sublattices being connected by a real space rotation symmetry \cite{vsmejkal2022emerging}. Altermagnet materials are posed to have a significant impact in the field of spintronics, as there is already a significant effort underway to replace ferromagnet-based technologies with antiferromagnets, which host faster magnetization dynamics and lack of stray field \cite{jungfleisch2018perspectives}. While the spin-degenerate band structure of antiferromagnets makes their N\'eel state difficult to detect, the spin-splitting in altermagnets is predicted to produce novel spin-transport phenomena that are N\'eel vector dependent, thus overcoming the challenges faced by traditional collinear antiferromagnets. Since the prediction of altermagnetism \cite{hayami2019momentum,yuan2020giant,vsmejkal2020crystal,mazin2021prediction}, the spintronics field has invested significant research effort on finding altermagnetic candidate materials, with none more popular than rutile RuO\textsubscript{2}. Although it had long been considered a paramagnet \cite{ryden1970magnetic}, in 2017 a report of neutron scattering provided the first indication suggesting the existence of collinear antiferromagnetism along the \textit{c}-axis with a small moment of 0.05~$\mu_{B}$ per Ru atom \cite{berlijn2017itinerant}. While many subsequent studies reported positive indications for altermagnetism in RuO\textsubscript{2} \cite{bai2022observation,bose2022tilted,feng2022anomalous,guo2024direct,karube2022observation,liao2024separation,liu2023inverse,bai2023efficient,fedchenko2024observation}, recent probes for the ordered antiferromagnetism by neutron scattering, muon spin rotation, and spin and angle-resolved photoelectron spectroscopy have suggested no ordered antiferromagnetism or altermagnetic band structure \cite{hiraishi2024nonmagnetic,kessler2024absence,liu2024absence}.

\begin{figure}[h]
\centering \includegraphics[width=1\columnwidth]{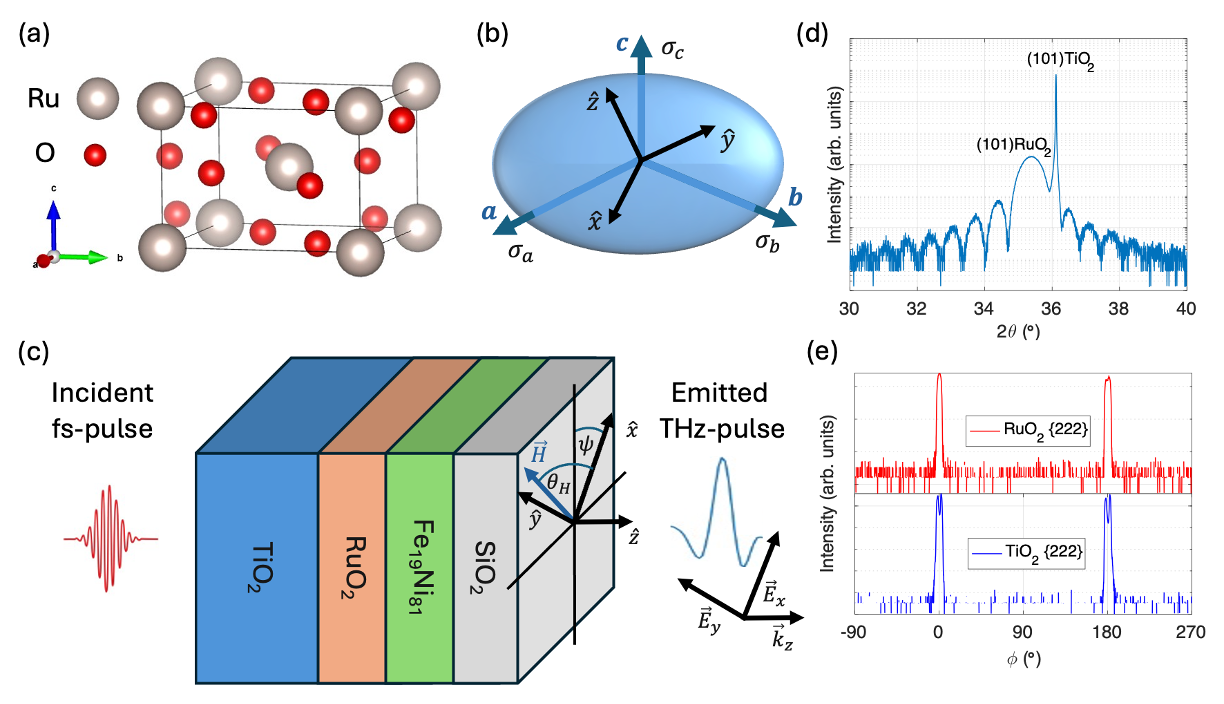}
\caption{(a) Rutile RuO\textsubscript{2} unit cell, (b) ellipsoidal conductivity tensor with the lab frame coordinates rotated relative to the crystal axes, (c) diagram of experimental setup for TDTS measurement with $\psi$ being the angle between the laboratory frame ($\hat{x}$,$\hat{y}$,$\hat{z}$) and the principle crystal axis, (d) HRXRD $\theta$/2$\theta$ measurement of (101) oriented sample and (e) in-plane $\phi$ scan of the asymmetric \{222\} planar family in both the film (red) and substrate (blue).}
\label{fig:1}
\end{figure}

Here, we investigate epitaxial RuO\textsubscript{2} in bilayer with ferromagnetic (FM) permalloy (Fe\textsubscript{19}Ni\textsubscript{81}) by time-domain terahertz (THz) spectroscopy (TDTS), probing for three possible mechanisms of laser-pulse (LP) induced charge dynamics: the relativistic inverse spin-Hall effect (ISHE), the non-relativistic inverse altermagnetic spin-splitting effect (IASSE), and the non-relativistic and non-magnetic electrical anisotropic conductivity (EAC). It is well known that FM/normal-metal(NM) bilayers can be excited by a femtosecond (fs)-LP to produce an ultrafast demagnetization process accompanied by charge dynamics that produce THz radiation, called a spintronic emitter (SE) \cite{wu2021principles}. 
The common interpretation is that the ultrafast demagnetization causes a spin-current ($\vec{J}_{s}$) to be injected from the FM into the NM, which then converts to a transverse charge current ($\vec{J}_{c}$) by the ISHE, following the equation $E_{THz}\propto\vec{J_{c}}=\theta_{SH}\vec{J_{s}}\times\vec{{\sigma}}$, where $\theta_{SH}$ is the spin Hall angle of the NM layer and $\vec{{\sigma}}$ is the spin-polarization. Recently, the IASSE  and EAC were reported for the first time as novel non-relativistic THz emission mechanisms, both in heterostructures comprising RuO\textsubscript{2} \cite{liu2023inverse,zhang2023nonrelativistic}. While the EAC mechanism only requires RuO\textsubscript{2} to be metallic and an epitaxial anisotropic crystal, the IASSE requires ordered antiferromagnetism and an altermagnetic band structure. The tetragonal unit cell and ellipsoid conductivity tensor are shown in Figure \ref{fig:1}(a)\&(b), in line with the most recent findings that RuO\textsubscript{2} is likely not an antiferromagnet. 
If RuO\textsubscript{2} is an altermagnet and not a NM, the IASSE mechanism should lead to spin-to-charge conversion that can contribute to the THz emission when the injected spin current $\vec{J_{s}}$ has a projection along the [100] direction and the spin-polarization $\vec{{\sigma}}$ has a projection along the N\'eel vector $\vec{N}$ (a more detailed explanation of the IASSE can be found in the supplementary information (SI) Figure S6). The EAC mechanism, however, does not involve either $\vec{J_{s}}$ or $\vec{N}$, but only requires a charge current $\vec{J_{c}}$ injected along a crystal direction which has off-diagonal conductivity tensor terms. In fact, in reference \cite{zhang2023nonrelativistic}, no FM layer is used and so no spin-current is related to their observed THz emission. The EAC mechanism is simply based on Ohm's law, $\vec{J_{c}}=\sigma\vec{E}$, where the conductivity tensor $\sigma$ is anisotropic because of the rutile crystal structure, with $\sigma_{a}=\sigma_{b}>\sigma_{c}$.  Orientations which contain components of both the \textit{a} and \textit{c}-axis lattice parameters projecting in the out-of-plane direction ($\hat{z}$) will thus show EAC based THz emission. In our work, we use rutile TiO\textsubscript{2} substrates to achieve epitaxial growth of the RuO\textsubscript{2} layer of four common orientations in order to study the crystal-direction dependent conversion: (001), (100), (110), and (101). Given that the altermagnetic band structure is a direct consequence of the real space crystal structure, it is advantageous to use a single-crystal or epitaxial thin-film to properly probe the crystal direction-dependent phenomena. An example of the high-quality epitaxial nature of our films, grown by reactive magnetron sputtering, is shown in Figure \ref{fig:1}(d)-(e). The high-resolution X-ray diffraction (HRXRD) $\theta/2\theta$ scan shows many Laue oscillations and the in-plane $\phi$ scan shows a single in-plane phase that is matched with the substrate, as is the case for all our samples which are reported in the SI Figures S1 \& S2. THz emission from the samples is measured by TDTS. A laser pulse incident from the substrate side pumps the spin-polarized charge current that is subsequently injected from the permalloy into the RuO\textsubscript{2} along the $-\hat{z}$ direction. The spin-polarization is controlled by the rotation of the applied magnetic field angle $\theta_{H}$, as depicted in Figure \ref{fig:1}(c). The \textit{x} and \textit{y} components of the emitted THz electric field are measured as a function of $\theta_{H}$ and integrated to find the root mean square value of the emitted amplitude, $E_{rms}$.

\section{Results}
\subsection{THz emission from non-EAC orientations of RuO\textsubscript{2}: (001), (100), and (110)}

\begin{figure}[h]
\centering \includegraphics[width=1\columnwidth]{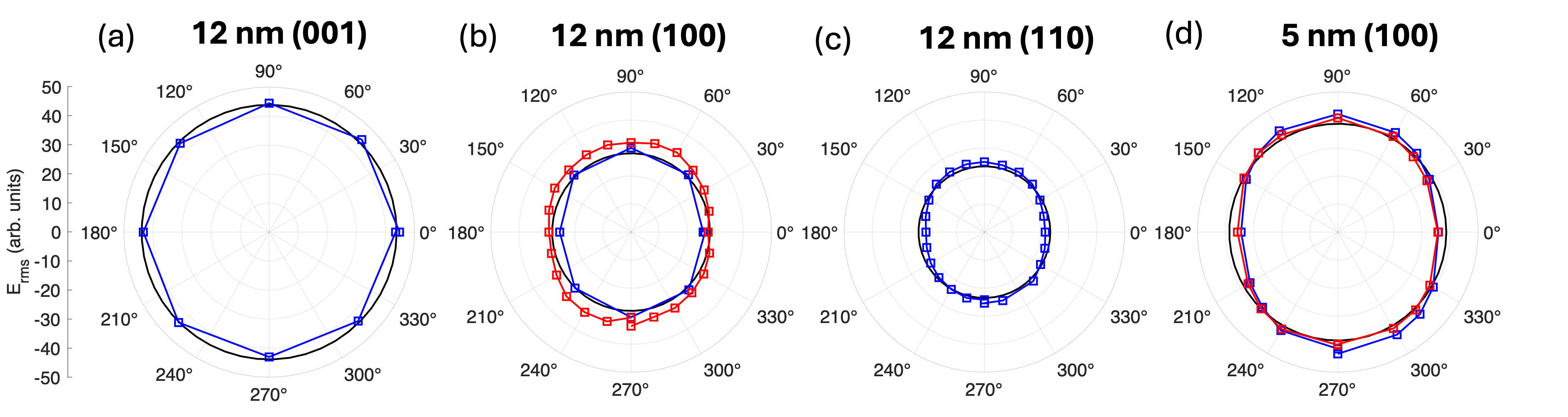}
\caption{\textbf{$E_{rms}$ as a function of external field angle $\theta_{H}$} for (a) (001), (b) (100), and (c) (110) samples with 12~nm thick RuO\textsubscript{2} layer, and (d) for (100) oriented sample with 5~nm thick RuO\textsubscript{2} layer. Blue data points are from as-deposited samples, red data points are post field-annealing, and black circles are guides to the eye (with a radius equal to the as-deposited $E_{rms}$ average).}
\label{fig:2}
\end{figure}

We first consider the samples with RuO\textsubscript{2} layer orientations that are not expected to have emission by EAC, i.e. the (001), (100), and (110) orientations, which have only the \textit{a}- or \textit{c}-axes in the out-of-plane direction, and thus have no off-diagonal conductivity tensor terms for a current injected along $\hat{z}$. Polar plots of the measured $E_{rms}$ under varied  $\theta_{H}$ are reported in Figure \ref{fig:2}. For the (001) sample in Figure  \ref{fig:2}(a), the emitted THz amplitude is clearly isotropic and well described by a circle, which is consistent with spin-to-charge conversion by the ISHE only, as expected for RuO\textsubscript{2} being either a NM or an altermagnet. For the (100) and (110) oriented samples in Figure \ref{fig:2}(b)-(d), the emitted THz amplitude is found to be slightly anisotropic, with the stronger emission occurring when the magnetization is parallel to the \textit{c}-axis of the crystals at $\theta_{H}=90^{\circ}$. We argue that the in-plane EAC of those samples causes the anisotropy, and not the IASSE. This matter is further discussed in the Discussion section. 

It is important to note that, in contrast to our results, the THz emission reported in reference \cite{liu2023inverse} that was attributed to the IASSE was only observed after field annealing of the samples. 
Additionally, their results indicated that the IASSE contribution to the emission varies strongly with thickness, with the strongest emission occurring for a 5~nm thick RuO\textsubscript{2} layer. In contrast, our results in Figure \ref{fig:2}(b)\&(d) show that the emission intensity (blue data points) is essentially the same for the (100) oriented samples with 12 and 5~nm thick RuO\textsubscript{2} layer. In addition, the emission is unchanged after field annealing following the same protocol used in reference \cite{liu2023inverse}, i.e. annealing at 475~K for 2 hours and subsequent cooling in an 8~kG field (red data points).

\subsection{THz emission from (101) oriented RuO\textsubscript{2}}

We now consider the (101) oriented sample. It has already been demonstrated that metallic (101) oriented rutile RuO\textsubscript{2} and IrO\textsubscript{2} generate THz emission by the non-magnetic and non-relativistic EAC mechanism that is enhanced by an additional NM layer \cite{zhang2023nonrelativistic}. 
However, the effect of the EAC mechanism on a spin-current that is injected from an FM layer into the (101) RuO\textsubscript{2} film has not yet been explored. 
Considering that both the IASSE and ISHE change sign when reversing the spin-polarization of the injected spin-current, but the EAC does not, we expect that the EAC contribution to the THz emission can be isolated by reversing the magnetic field direction.
Adding the THz emission collected at $\theta_{H}$ and $\theta_{H}+180^\circ$ and dividing by two (Equation \ref{eq:E_EAC}) results in the EAC contribution that does not depend on the field direction. 

\begin{equation}
E_{x,y,EAC} =(E_{x,y}(\theta_{H})+E_{x,y}(\theta_{H}+180^\circ))/2
\label{eq:E_EAC}
\end{equation}

\begin{figure}[h]
\centering \includegraphics[width=1\columnwidth]{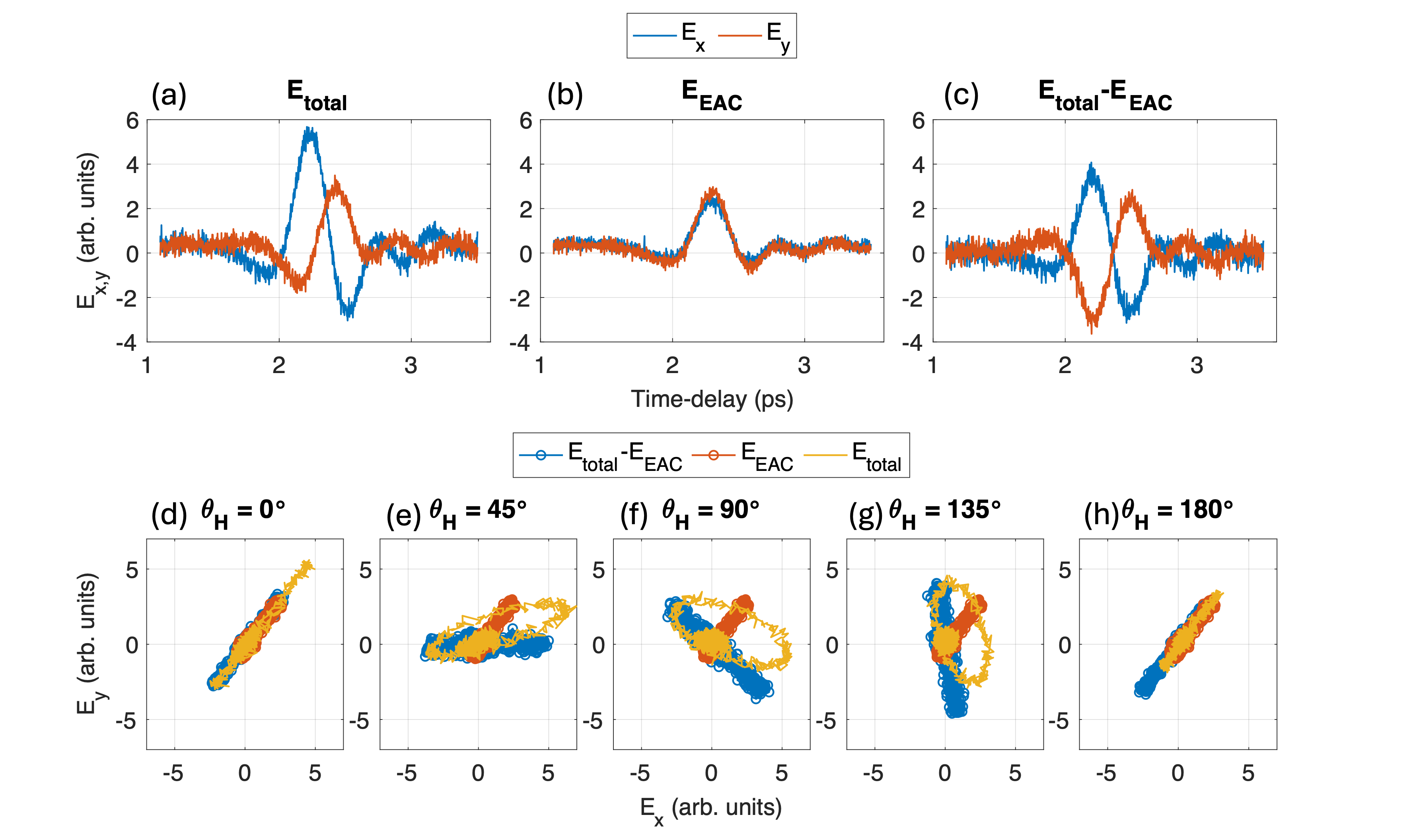}
\caption{Time-domain traces of the emitted THz wave electric field components of (101) oriented sample with 12 nm thick RuO\textsubscript{2} layer at $\psi=0^{\circ}$ and $\theta_{H}=90^{\circ}$ (a) total electric field, (b) EAC component, and (c) total electric field with EAC component subtracted. (d)-(h) Parametric plots of E\textsubscript{x} vs E\textsubscript{y} for field angles between $\theta_{H}=0^{\circ}$ and $180^{\circ}$, with total signal (yellow), EAC component (red), and total signal minus EAC component (blue).}
\label{fig:3}
\end{figure}

The \textit{x} and \textit{y} components of the total measured signal for the (101) sample at crystal angle $\psi=0^{\circ}$ (see Figure \ref{fig:1}(c)) and field angle $\theta_{H}=90^{\circ}$ are plotted in the time-domain in Figure \ref{fig:3}(a), the extracted EAC component of the signal is plotted in Figure \ref{fig:3}(b), and the total signal remaining after subtracting the EAC component is shown in Figure \ref{fig:3}(c). 
In Figure \ref{fig:3}(d)-(h), parametric plots of the \textit{x} and \textit{y} components of the emitted electric field for the signal excluding the EAC component is shown in blue, the EAC component is shown in red, and the total THz emission is shown in yellow. 
It can clearly be seen, that the combination of the two emission mechanisms (ISHE and EAC) can result in elliptically polarized light depending on the direction of the external magnetic field (e.g. yellow for $\theta_{H}=90^{\circ}$).

\begin{figure}[h]
\centering \includegraphics[width=1\columnwidth]{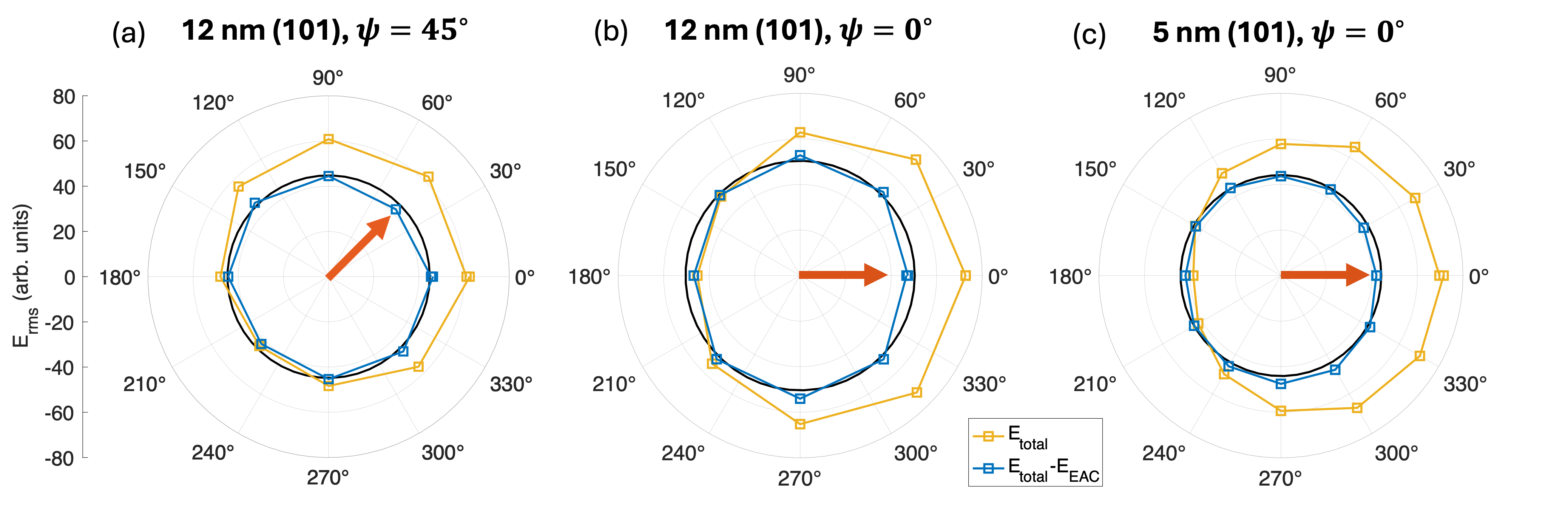}
\caption{E\textsubscript{rms} of the (101) samples. (a) The 12 nm sample at $\psi=45^{\circ}$, and (b) at $\psi=0^{\circ}$. (c) The 5 nm thick RuO\textsubscript{2} sample at $\psi=0^{\circ}$. The red arrows indicate the direction of the EAC emission.}
\label{fig:4}
\end{figure}

Figure \ref{fig:4} shows a polar plot of the $E_{rms}$ of the total emitted THz signal in yellow and the signal excluding the EAC component in blue. Figure \ref{fig:4}(a) \& (b) illustrate that the EAC component follows the rotation of the crystal by the angle $\psi$ (red arrow), as expected. The signal excluding EAC (blue) shows the same small anisotropy that was observed for the (100) and (110) samples (Fig. \ref{fig:2} b-d) that will be discussed below.

%We again analyze  $E_{rms}$ of the total emitted THz wave, and also the $E_{rms}$ after subtraction of the EAC component $E_{EAC}$. The results are shown in Figure \ref{fig:4} with the total emission in red and the emission without EAC in yellow. Figure \ref{fig:4}(a)\&(b) illustrate how the EAC component of emission rotates with the crystal angle $\psi$, and Figure \ref{fig:4}(c) shows that the elliptical nature persists when the RuO\textsubscript{2} layer is 5 nm thick. In the sample with 12 nm thick RuO\textsubscript{2} layer, a small anisotropy is present after subtracting  $E_{EAC}$, similar to the (100) and (110) samples. 

\section*{Discussion}

\textbf{Anisotropy of THz emission from (100), (110) and (101) orientations}

\begin{figure}[h]
\centering \includegraphics[width=1\columnwidth]{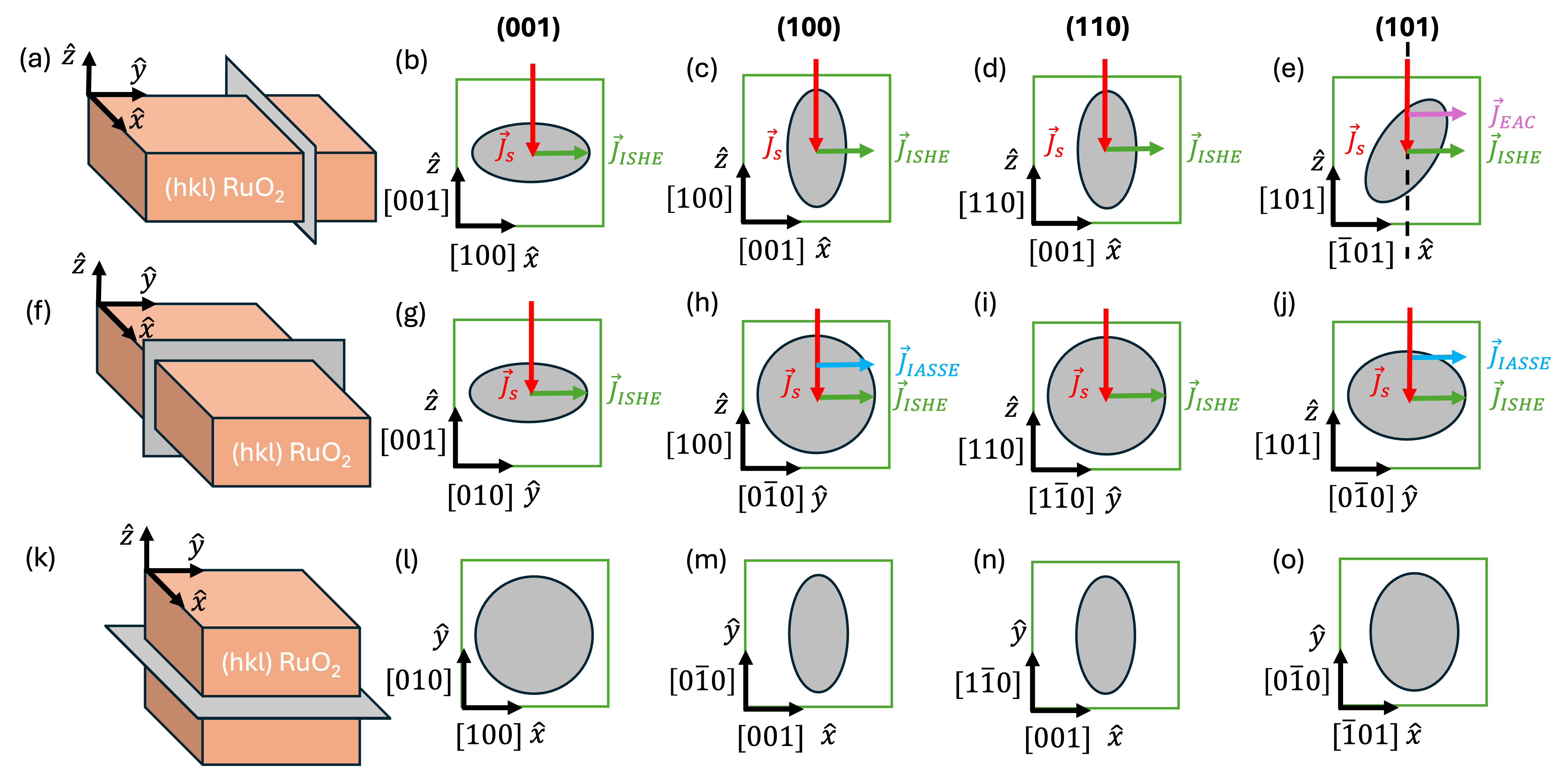}
\caption{{\textbf{2D projection of the ellipsoidal conductivity tensor for each sample orientation}}. Panels (a), (f), and (k) illustrate the plane of the sample examined for each row of projections. Panels (b)-(e) are the \textit{x-z} plane, (g)-(j) the \textit{y-z} plane, and (l)-(o) the \textit{x-y} plane. In panels containing \textit{z} direction, the pumped spin current $\vec{J}_{sp}$ is polarized into the plane, and the three possible THz emission mechanisms (ISHE, IASSE, and EAC) are included with their appropriate crystal direction dependencies.}
\label{fig5:EAC}
\end{figure}

We postulate that the small anisotropy of the THz emission observed in the (100), (110), and (101) oriented samples is due to in-plane EAC of RuO\textsubscript{2}, which modulates the in-plane charge currents that result from the ISHE.
%give rise to the THz emission. 
To illustrate the out-of-plane EAC mechanism of THz emission, and the in-plane EAC effect on THz emission from ISHE, the 2D projections of the conductivity tensor are shown in Figure \ref{fig5:EAC}. 
Panels (a)-(e) show the \textit{x-z} planes of each (hkl) orientation, (f)-(j) show the \textit{y-z} plane, and  (k)-(o) shows the \textit{x-y} plane. 
Since the (100), (110) and (101) oriented films contain both the \textit{a} and \textit{c}-axis of the lattice in their plane, the in-plane projection of the conductivity tensor is elliptical. 
When the external field is rotated, the in-plane charge current produced by the ISHE experiences different conductivities depending on the crystal directions; the amplitude of the current decays faster in the less conductive $\sigma_{c}$ direction. 
While we do not measure the direction-dependent in-plane conductivity of our samples, it is known that the rutile crystal of RuO\textsubscript{2} shows anisotropic conductance.
Since we observe EAC-induced THz emission from our (101) RuO\textsubscript{2} sample, 
EAC is expected to exist in RuO\textsubscript{2} samples with other orientations.

To exclude that the small anisotropy is caused by magnetic anisotropy in the FM layer we measured magnetization hysteresis loops of each sample along both flat edge directions by vibrating sample magnetometry (VSM), and report the results in the SI (Figures S3-S5). We observe an easy and hard axis in each sample that showed anisotropy in the THz emission amplitude, however the hard axis anisotropy field H\textsubscript{a} is less than 1 kG for any given sample, so it is reasonable to expect that all samples were saturated by the external applied field with a field strenth of 1.4 kG. The high symmetry (001) sample had an easy-plane magnetic anisotropy.

\textbf{Absence of IASSE in RuO\textsubscript{2}} 

If RuO\textsubscript{2} was an altermagnet, we would expect to observe THz emission from the IASSE in the (100) and (101) films.
The linear polarization of the IASSE-induced emission should follow the rotation of the external magnetic field that aligns $\vec{\sigma}$ relative to the N\'eel vector. 
%field that strong variation in the emission as the field rotates $\vec{\sigma}$ relative to the N\'eel vector. 
However, our measurements can be explained by a combination of ISHE, out-of-plane, and in-plane EAC. 
As discussed above, the small anisotropy can be explained by in-plane EAC, if it were induced by IASSE it should be absent for the (110) sample, for which the IASSE is forbidden.
%Although we do observe anisotropy in the THz emission amplitude of both the (100) and (101) films (after subtraction of EAC component in the (101) case), several features of our results suggest that the source is not IASSE. First of all, we observe a similar anisotropy in the emission of the (110) sample, for which the IASSE is forbidden. 
%Second, the amplitude anisotropy we observe is very small, although the IASSE is a non-relativistic effect which is expected to be much stronger than the relativistic ISHE. 
In addition, the THz emission anisotropy is essentially the same for both as-prepared and field-annealed (100) samples (Figure \ref{fig:2}(b)\&(d)). However, one would expect field annealing above the N\'eel temperature of $\approx400 K$ would align the antiferromagnetic domains and enhance the IASSE signal. 
These results clearly demonstrate the absence of the IASSE in our RuO\textsubscript{2} films. Our findings align with recent observations from neutron and muon spin rotation experiments \cite{hiraishi2024nonmagnetic,kessler2024absence}. Notably, all our samples were epitaxially grown on single crystal TiO\textsubscript{2} substrates, whereas in reference \cite{liu2023inverse}, the use of cubic MgO and YSZ substrates are known to result in multiple in-plane crystal phases, which could be related to the reported anisotropy. Our results prompt a reexamination of theoretical models and suggest further research to determine whether defects and stoichiometry of RuO\textsubscript{2} films may account for the varying experimental outcomes regarding the presence or absence of IASSE in RuO\textsubscript{2}.

\textbf{Implications of EAC and ISHE combined elliptical THz emitter}

We emphasize that the results from the (101) oriented sample give rise to a novel method for producing elliptical polarized THz light due to the superposition of THz radiation from two sources (ISHE and EAC).
The tunability of the polarization direction of THz from ISHE by an external magnetic field gives control over the ellipticity and chirality of the resulting field.
In addition, unlike the ISHE, which predominantly occurs at interfaces within the spin diffusion length of RuO\textsubscript{2} the emission from the EAC effect is a bulk phenomenon. Thus, by adjusting the RuO\textsubscript{2} layer thickness, it is possible to control the relative phase between the two emitters, offering flexible options for pulse shaping. 
Although recently, a hybrid THz emitter consisting of a photo-conductive antenna (PCA) and SE have been combined to produce elliptical polarized THz \cite{wu2024hybrid}, the current method is more straightforward and can be easily implemented without the need for patterning a PCA structure. 
Other methods to achieve polarization and pulse shape control of THz waves require modulation and filtering, which limits other wave properties such as bandwidth \cite{sato2013terahertz}. The EAC and ISHE both produce essentially the same bandwidth of THz frequencies, whereas the combination of PCA and SE hybrid emitter  relies on the PCA component, which is limited to approximately 3 THz bandwidth. 
Our elliptical emitter can be optimized by using an EAC material with higher spin-orbit coupling (SOC), or otherwise higher $\beta$ parameter as described by Zhang et al. in reference \cite{zhang2023nonrelativistic}, such as using IrO\textsubscript{2} instead of RuO\textsubscript{2}. Addition of a high-SOC NM layer in-between the EAC and FM layers could also improve emission strength by enhancing the ISHE signal from the FM and also possibly pumping more charge current into the EAC layer.

\section*{Conclusion}

We have studied the laser-pulse induced charge dynamics in high quality epitaxial RuO\textsubscript{2} and permalloy bilayers by TDTS. Our results reveal no evidence of the IASSE signature of altermagnetism in either the (100) or (101) orientations for which it is expected, including after field annealing of the (100) samples. However, we observed the non-magnetic and non-relativistic EAC emission in the (101) oriented sample. We also observe for the first time an angle-dependent THz amplitude in the (100), (110), and (101) oriented samples that we assign to in-plane modulation of the ISHE converted charge current by EAC. We demonstrated that the combination of ISHE and EAC THz emission mechanisms in the (101) oriented sample can generate elliptically polarized THz emission with chirality modulated by the applied magnetic field. Our findings imply that RuO\textsubscript{2} is a normal metal and not an altermagnet, and establish a novel, easily implemented approach for generating tunable, elliptically polarized THz light source using an external field.  

\newpage
\section*{Methods}

\textbf{Sample fabrication} 

RuO\textsubscript{2} films were deposited on double side polished TiO\textsubscript{2} substrates by reactive magnetron sputtering in a vacuum chamber with base pressure of 5x$10^{-8}$ Torr. Substrates were heated to 600 $^{\circ}$C and baked for 10 min prior to deposition. The Ru target was sputtered with 200 W RF power in an atmosphere of 5:1 Ar/O\textsubscript{2} gas mixture with a total pressure of 15 mTorr. Subsequent layers of 8 nm thick permalloy (Fe\textsubscript{19}Ni\textsubscript{81}) and 10 nm thick SiO\textsubscript{2} capping layer were deposited in-situ after cooling substrates to room temperature, using 5 mTorr total pressure of Ar and RF bias of 100 and 200 W, respectively. The permalloy target was pre-sputtered for 10 min to clean any oxide from the target surface.

\textbf{High-resolution X-ray diffraction}

X-ray diffraction patterns were measured using a Rigaku Smartlab diffractometer with a Ge (220) double bounce monochromator.

\textbf{Vibrating sample magnetometry}

A Quantum-Design VersaLab vibrating sample magnetometer was used to measure room temperature hysteresis loops with an in-plane magnetic field. Field annealing of the (100) samples was performed in the VSM using an oven sample holder.

\textbf{Time-domain terahertz spectroscopy}

Samples were illuminated from the substrate side (to eliminate the substrate birefringence effect on the polarization of the emitted THz wave) with a laser pulse of approximately 40 fs in pulse duration and a center wavelength of 800 nm at 10 kHz. The emitted THz wave was measured using electro-optic sampling with a 1~mm thick (110)-ZnTe crystal. The sample was placed in a constant field of 1.4 kG provided by a pair of permanent magnets, which were rotated relative to the sample. A pair of THz polarizers was used to extract the total electric field of the THz radiation.

\newpage
\section*{Author contribution}
D.T.P. and J.Q.X. conceived the project. D.T.P. and N.J.P. deposited the films and characterized them by HRXRD and VSM, and also field annealed the samples. S.B. and X.W. deposited samples that were used for preliminary TDTS measurements (by W.W.) that resolved the substrate birefringence issue via backside incidence excitation. L.S., S.S., and W.W. performed the TDTS measurements with supervision by L.G. and M.B.J. Data analysis was done by D.T.P. with assistance from L.S. and N.J.P. All authors discussed the results and provided feedback.

\section*{Competing interests}
D.T.P., L.S., S.S., X.W., W.W., L.G., M.B.J., and J.Q.X are listed as inventors on a pending patent application by the University of Delaware related to observations in this manuscript. The remaining authors declare no competing interest.

\section*{Data availability}
All data is available upon request from dplouff@udel.edu.

\section*{Acknowledgments}
This research was sponsored by NSF DMR-1904076, NSF through the University of Delaware Materials Research Science and Engineering Center (MRSEC), DMR-2011824, and King Abdullah University of Science and Technology (KAUST), ORFS-2022-CRG11-5031.2.  

\newpage
\section*{Supplemental Information}

\setcounter{figure}{0}
\renewcommand{\figurename}{FIG.}
\renewcommand{\thefigure}{S\arabic{figure}}

\textbf{Structural characterization of samples}

The wide range $\theta$/$2\theta$ scans of each sample with 12 nm thick RuO\textsubscript{2} layer are shown in Figure S1. For each substrate peak there is a reflection of the same planes for the film indicating a single out-of-plane crystal phase. All samples show Laue oscillations for the lowest index peak reflections, and the (110) and (101) peaks in particular have many strong oscillations indicating a high degree of crystal quality. On the (100) and (110) oriented films a small (111) peak from the permalloy layer is also clearly present, while in the (101) sample is peaking out of the noise and in the (001) sample there is no peak at all.

We also find that there is only one in-plane crystal phase of the film which is also oriented with the substrate as indicated by the identical symmetry of the asymmetric peaks examined by HRXRD $\phi$ scan in Figure S2. In Figure S2(d)\&(h), a $\phi$ scan of the asymmetric {103} peak in the (101) sample shows only one peak, illustrating the lack of mirror symmetry which is responsible for the EAC mechanism of THz emission.
\begin{figure}[h]
\centering \includegraphics[width=1\columnwidth]{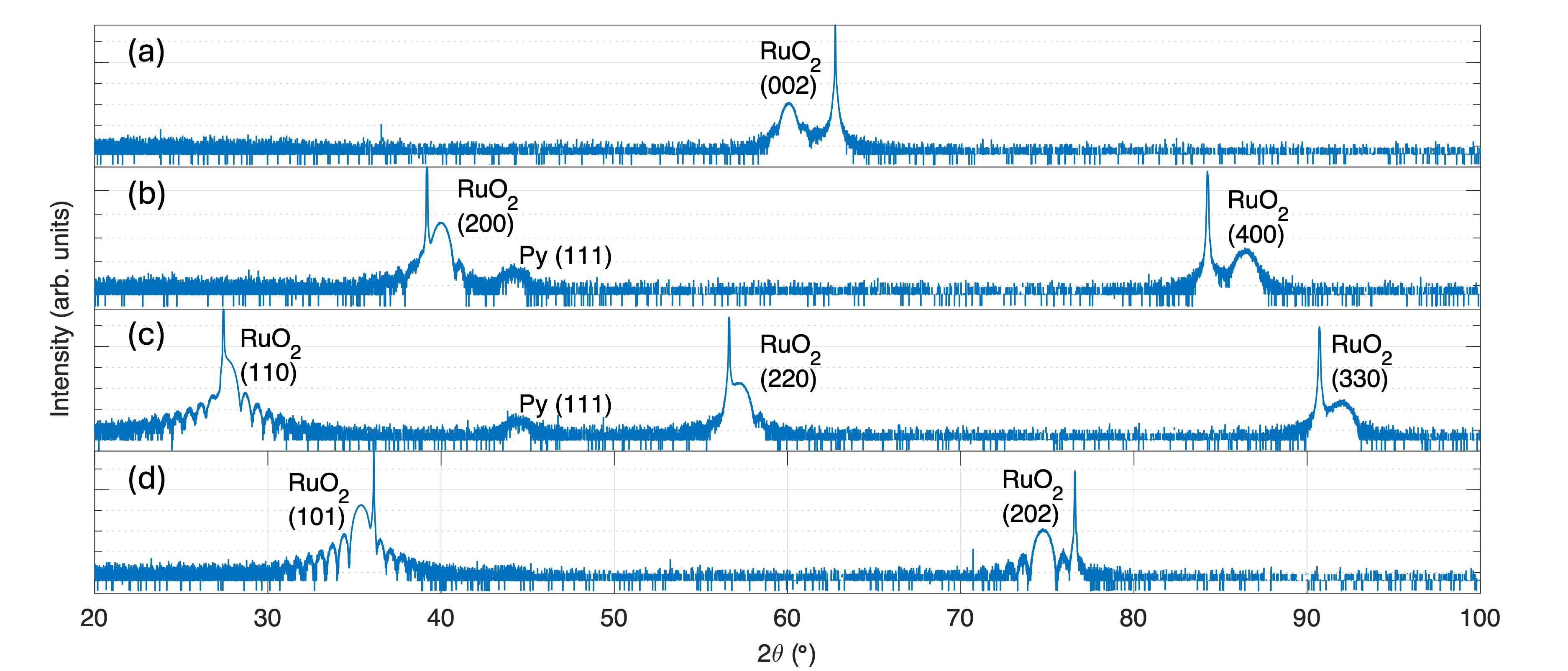}
\caption{{\textbf{Out-of-plane structural characterization}}. Wide-range $\theta$/$2\theta$ scans of each sample with 12 nm thick RuO\textsubscript{2} layer of each orientation (a) (001), (b) (100), (c) (110), and (d) (101).}
\label{supfig:gonios}
\end{figure}
\newpage

\begin{figure}[h]
\centering \includegraphics[width=1\columnwidth]{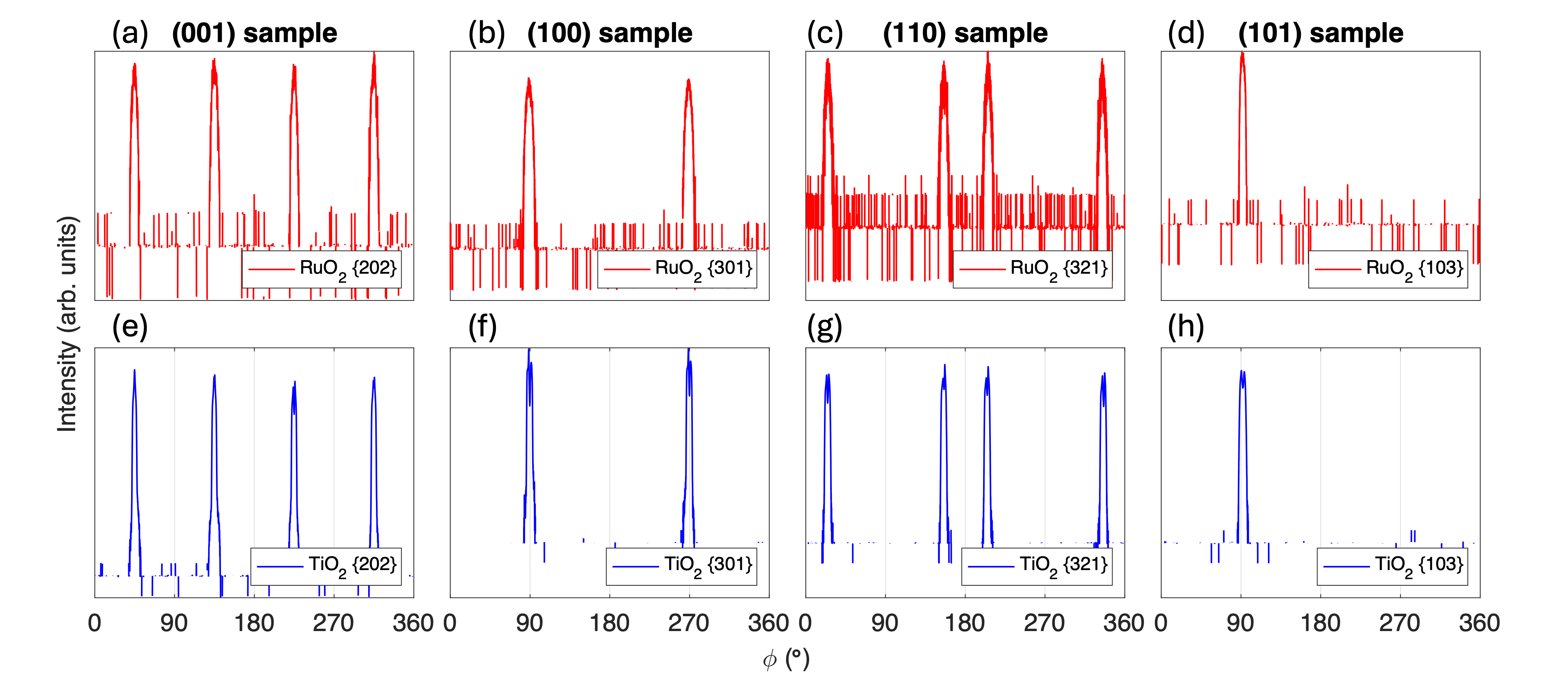}
\caption{{\textbf{In-plane structural characterization}}. Full-range $\phi$ scans of each sample with 12 nm thick RuO\textsubscript{2} layer of each orientation: (a) \& (e) (001), (b) \& (f) (100), (c) \& (g) (110), and (d) \& (h) (101). Top panels (a)-(d) are film, and bottom panels (e)-(h) are substrate.}
\label{supfig:phis}
\end{figure}

\textbf{Magnetic properties of the permalloy layer}

Magnetization hysteresis loops were measured along both flat edges of each sample using VSM. For samples of each orientation with 12 nm thick RuO\textsubscript{2} layer, the results are reported in Figure S3, and the (100) and (101) samples with 5 nm thick RuO\textsubscript{2} layer are reported in Figure S4. The (001) oriented sample shows two equivalent easy axes, while the other orientations each show an easy magnetic axis along the \textit{c}-axis direction. The coercivity (H\textsubscript{c}) and anisotropy field (H\textsubscript{a}) values of each sample varies and is reported in Table S1. Both (100) samples were field annealed following the same protocol as Ref. \cite{liu2023inverse}, and were then remeasured at room-temperature. The hysteresis loops are shown in Figure S5, and H\textsubscript{c} and H\textsubscript{a} are reported in Table S2, demonstrating the magnetic properties are not changed in any substantial way by the annealing. We speculate that the variation of H\textsubscript{c} and H\textsubscript{a} with the crystal orientation of the RuO\textsubscript{2} layer is related to the symmetry of the interface, and is not caused by an exchange-coupling phenomena which would require antiferromagnetism in RuO\textsubscript{2}. 

\begin{figure}[h]
\centering \includegraphics[width=0.9\columnwidth]{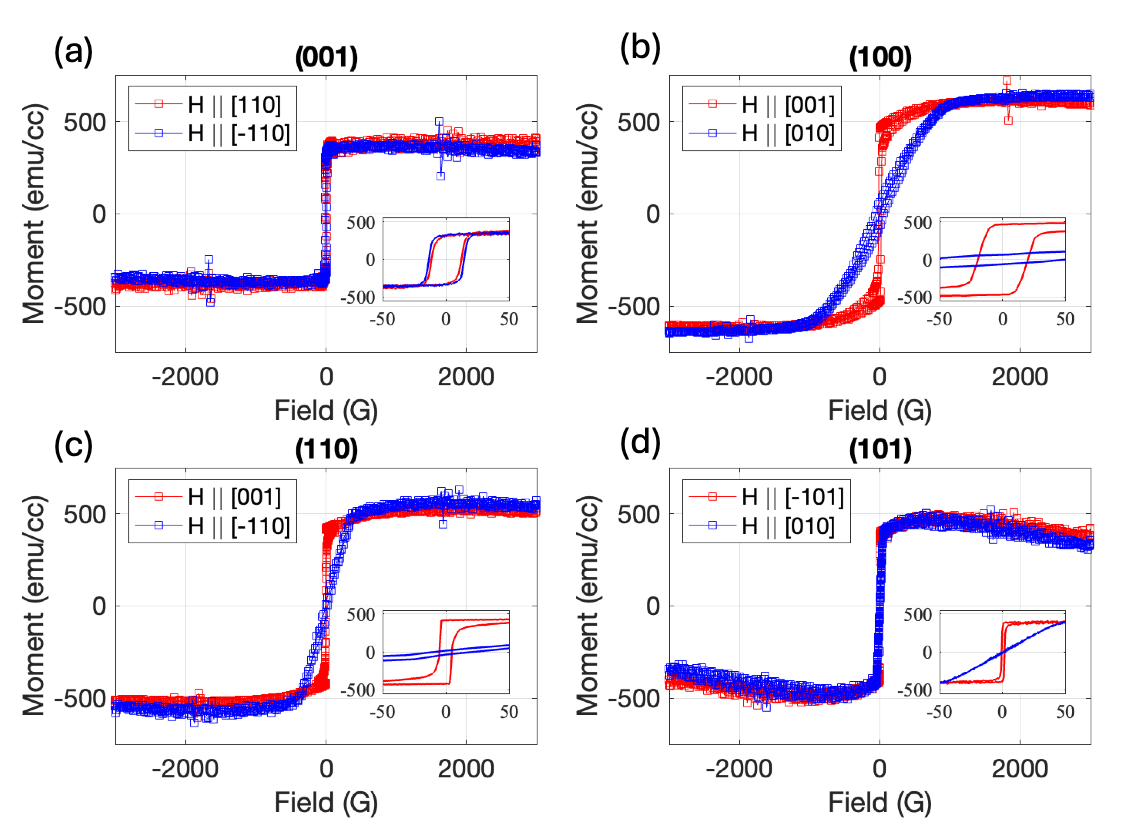}
\caption{{\textbf{VSM}}. Magnetization hysteresis loops of samples with 12 nm thick RuO\textsubscript{2} layer measured along both flat edge directions. Inset figures show the same figure zoomed in at small field values. Panels show each orientation of RuO\textsubscript{2} layer: (a) (001), (b) (100), (c) (110), and (d) (101). }
\label{supfig:VSM_all}
\end{figure}

\begin{figure}[h]
\centering \includegraphics[width=1\columnwidth]{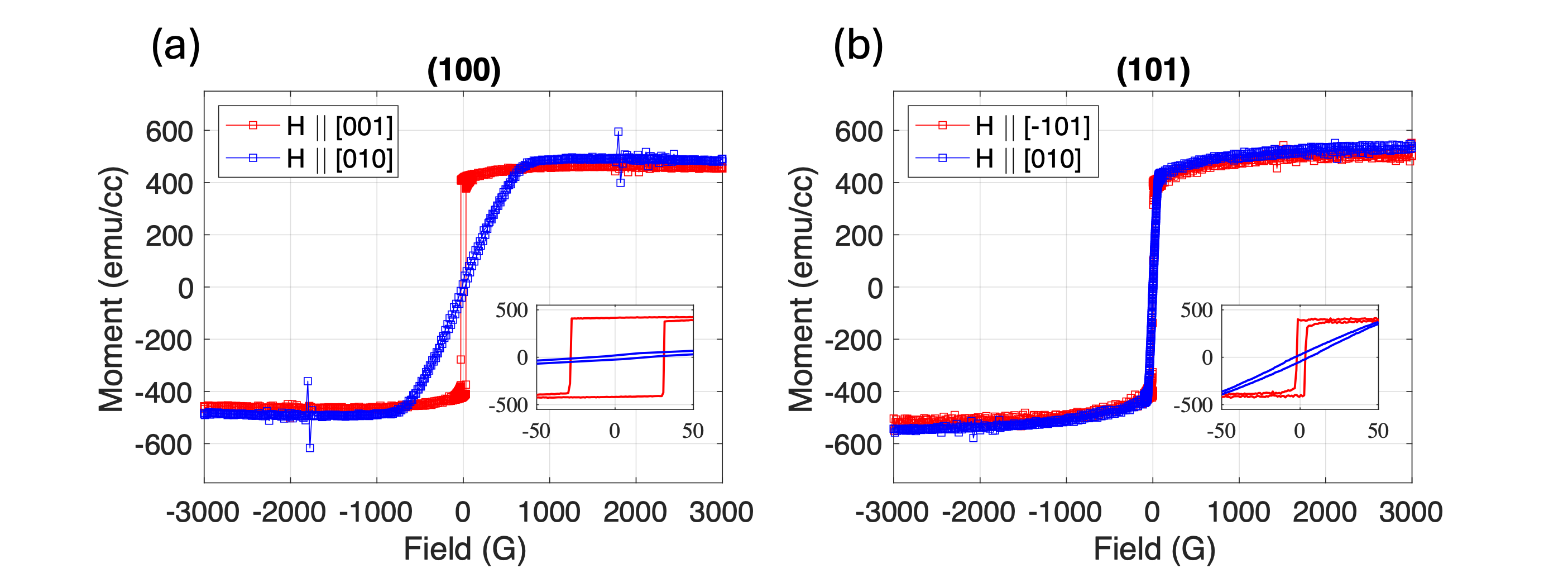}
\caption{{\textbf{VSM}}. Magnetization hysteresis loops for samples with 5 nm thick RuO\textsubscript{2} layer, measured along both flat edge directions.}
\label{supfig:VSM_5nm}
\end{figure}

\begin{table}
\begin{center}
\caption{Magnetic properties of as-deposited samples.}
\begin{tabular}{||c c c c||} 
 \hline
 (hkl) & RuO\textsubscript{2} thickness (nm) & H\textsubscript{c} (G)& H\textsubscript{a} (G)\\ [0.5ex] 
 \hline\hline
 (001) & 12 & 12 & 20 \\ 
 \hline
 (100) & 12 & 20 & 1000 \\
 \hline
 (110) & 12 & 5 & 400 \\ 
 \hline
  (101) & 12 & 2 & 50 \\
 \hline
  (100) & 5 & 30 & 800 \\
 \hline
  (101) & 5 & 2 & 60 \\ %[1ex] 
 \hline
\end{tabular}
\end{center}
\end{table}

\begin{figure}[h]
\centering \includegraphics[width=1\columnwidth]{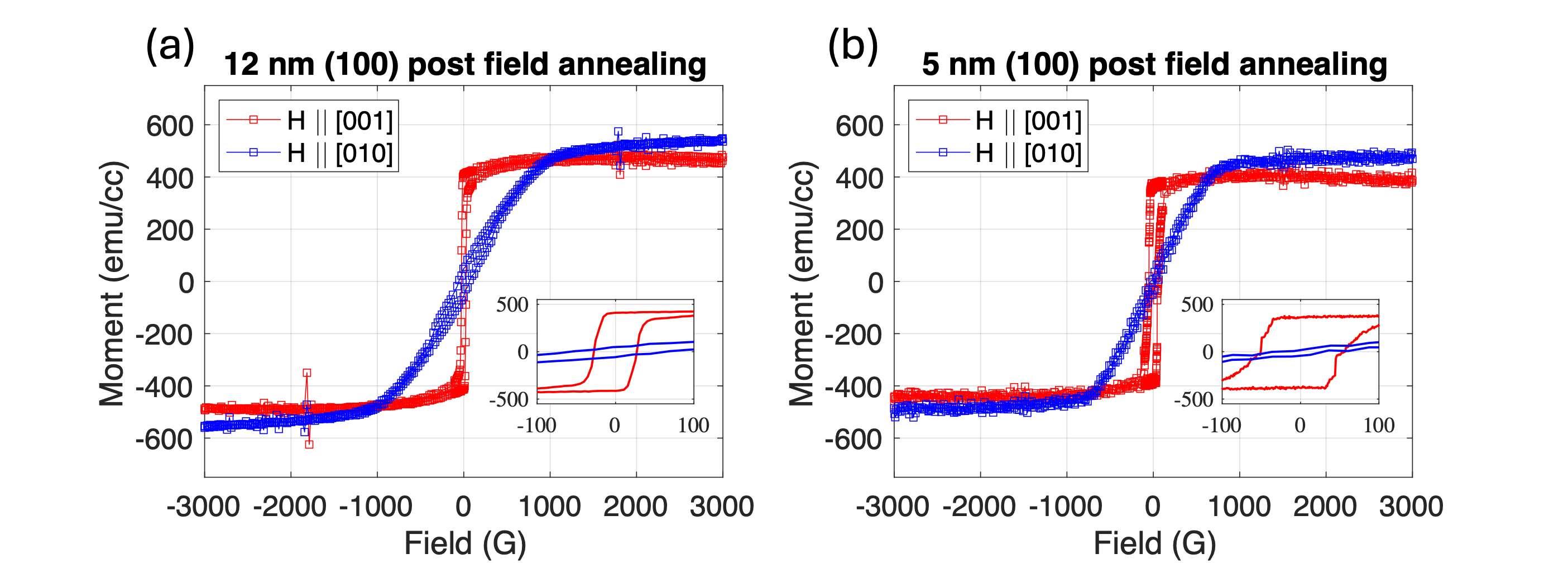}
\caption{{\textbf{VSM post field annealing}}. (a) 12 nm thick RuO\textsubscript{2} layer, (b) 5 nm thick.}
\label{supfig:VSM_posttrain}
\end{figure}

\begin{table}
\begin{center}
\caption{Magnetic properties of field annealed samples.}
\begin{tabular}{||c c c c||} 
 \hline
 (hkl) & RuO\textsubscript{2} thickness (nm) & H\textsubscript{c} (G)& H\textsubscript{a} (G)\\ [0.5ex] 
 \hline\hline
 (100) & 12 & 25 & 1000 \\
 \hline
  (100) & 5 & 50 & 800 \\
 \hline
\end{tabular}
\end{center}
\end{table}

In Figure S6(a), the rutile RuO\textsubscript{2} unit cell is shown with presumed antiferromagnetism along the c-axis, while in Figure S6(b) the altermagnetic spin-split Fermi surface illustrates the IASSE. When a spin current is injected along the [100] direction, the anisotropic conductivity of the two spin populations gives rise to a transverse charge current, so long as the spin polarization is projected on the N\'eel vector. In the supplement of reference \cite{liu2023inverse}, it was suggested that $\vec{J}_c\propto\vec{J}_s\times\vec{N}$, however, this is unlikely as the spin polarization is not involved. Given that the anisotropic spin conductivities are only for up or down spins along the \textit{c} direction, a spin current injected with any other spin polarization would not be affected by the spin-split Fermi surface and thus would not contribute to IASSE. 

\begin{figure}[h]
\centering \includegraphics[width=0.75\columnwidth]{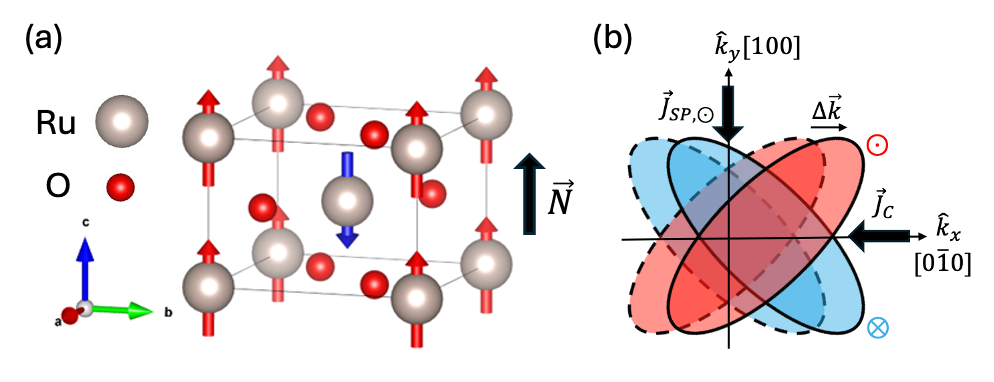}
\caption{(a) RuO\textsubscript{2} unit cell presuming c-axis antiferromagnetism, (b) (001) Fermi surface cut presuming d-wave altermagnetism with an injected spin current $\vec{J_{s}}$ that is converted to transverse charge current $\vec{J_{c}}$ by the IASSE.}
\label{fig:AFM_FermiSurf}
\end{figure}

%\nocite{*}
\providecommand{\noopsort}[1]{}\providecommand{\singleletter}[1]{#1}%


\providecommand{\noopsort}[1]{}\providecommand{\singleletter}[1]{#1}%
\begin{thebibliography}{24}%
\makeatletter
\providecommand \@ifxundefined [1]{%
 \@ifx{#1\undefined}
}%
\providecommand \@ifnum [1]{%
 \ifnum #1\expandafter \@firstoftwo
 \else \expandafter \@secondoftwo
 \fi
}%
\providecommand \@ifx [1]{%
 \ifx #1\expandafter \@firstoftwo
 \else \expandafter \@secondoftwo
 \fi
}%
\providecommand \natexlab [1]{#1}%
\providecommand \enquote  [1]{``#1''}%
\providecommand \bibnamefont  [1]{#1}%
\providecommand \bibfnamefont [1]{#1}%
\providecommand \citenamefont [1]{#1}%
\providecommand \href@noop [0]{\@secondoftwo}%
\providecommand \href [0]{\begingroup \@sanitize@url \@href}%
\providecommand \@href[1]{\@@startlink{#1}\@@href}%
\providecommand \@@href[1]{\endgroup#1\@@endlink}%
\providecommand \@sanitize@url [0]{\catcode `\\12\catcode `\$12\catcode `\&12\catcode `\#12\catcode `\^12\catcode `\_12\catcode `\%12\relax}%
\providecommand \@@startlink[1]{}%
\providecommand \@@endlink[0]{}%
\providecommand \url  [0]{\begingroup\@sanitize@url \@url }%
\providecommand \@url [1]{\endgroup\@href {#1}{\urlprefix }}%
\providecommand \urlprefix  [0]{URL }%
\providecommand \Eprint [0]{\href }%
\providecommand \doibase [0]{https://doi.org/}%
\providecommand \selectlanguage [0]{\@gobble}%
\providecommand \bibinfo  [0]{\@secondoftwo}%
\providecommand \bibfield  [0]{\@secondoftwo}%
\providecommand \translation [1]{[#1]}%
\providecommand \BibitemOpen [0]{}%
\providecommand \bibitemStop [0]{}%
\providecommand \bibitemNoStop [0]{.\EOS\space}%
\providecommand \EOS [0]{\spacefactor3000\relax}%
\providecommand \BibitemShut  [1]{\csname bibitem#1\endcsname}%
\let\auto@bib@innerbib\@empty
%</preamble>
\bibitem [{\citenamefont {{\v{S}}mejkal}\ \emph {et~al.}(2022)\citenamefont {{\v{S}}mejkal}, \citenamefont {Sinova},\ and\ \citenamefont {Jungwirth}}]{vsmejkal2022emerging}%
  \BibitemOpen
  \bibfield  {author} {\bibinfo {author} {\bibfnamefont {L.}~\bibnamefont {{\v{S}}mejkal}}, \bibinfo {author} {\bibfnamefont {J.}~\bibnamefont {Sinova}},\ and\ \bibinfo {author} {\bibfnamefont {T.}~\bibnamefont {Jungwirth}},\ }\bibfield  {title} {\bibinfo {title} {Emerging research landscape of altermagnetism},\ }\href@noop {} {\bibfield  {journal} {\bibinfo  {journal} {Physical Review X}\ }\textbf {\bibinfo {volume} {12}},\ \bibinfo {pages} {040501} (\bibinfo {year} {2022})}\BibitemShut {NoStop}%
\bibitem [{\citenamefont {Jungfleisch}\ \emph {et~al.}(2018)\citenamefont {Jungfleisch}, \citenamefont {Zhang},\ and\ \citenamefont {Hoffmann}}]{jungfleisch2018perspectives}%
  \BibitemOpen
  \bibfield  {author} {\bibinfo {author} {\bibfnamefont {M.~B.}\ \bibnamefont {Jungfleisch}}, \bibinfo {author} {\bibfnamefont {W.}~\bibnamefont {Zhang}},\ and\ \bibinfo {author} {\bibfnamefont {A.}~\bibnamefont {Hoffmann}},\ }\bibfield  {title} {\bibinfo {title} {Perspectives of antiferromagnetic spintronics},\ }\href@noop {} {\bibfield  {journal} {\bibinfo  {journal} {Physics Letters A}\ }\textbf {\bibinfo {volume} {382}},\ \bibinfo {pages} {865} (\bibinfo {year} {2018})}\BibitemShut {NoStop}%
\bibitem [{\citenamefont {Hayami}\ \emph {et~al.}(2019)\citenamefont {Hayami}, \citenamefont {Yanagi},\ and\ \citenamefont {Kusunose}}]{hayami2019momentum}%
  \BibitemOpen
  \bibfield  {author} {\bibinfo {author} {\bibfnamefont {S.}~\bibnamefont {Hayami}}, \bibinfo {author} {\bibfnamefont {Y.}~\bibnamefont {Yanagi}},\ and\ \bibinfo {author} {\bibfnamefont {H.}~\bibnamefont {Kusunose}},\ }\bibfield  {title} {\bibinfo {title} {Momentum-dependent spin splitting by collinear antiferromagnetic ordering},\ }\href@noop {} {\bibfield  {journal} {\bibinfo  {journal} {journal of the physical society of japan}\ }\textbf {\bibinfo {volume} {88}},\ \bibinfo {pages} {123702} (\bibinfo {year} {2019})}\BibitemShut {NoStop}%
\bibitem [{\citenamefont {Yuan}\ \emph {et~al.}(2020)\citenamefont {Yuan}, \citenamefont {Wang}, \citenamefont {Luo}, \citenamefont {Rashba},\ and\ \citenamefont {Zunger}}]{yuan2020giant}%
  \BibitemOpen
  \bibfield  {author} {\bibinfo {author} {\bibfnamefont {L.-D.}\ \bibnamefont {Yuan}}, \bibinfo {author} {\bibfnamefont {Z.}~\bibnamefont {Wang}}, \bibinfo {author} {\bibfnamefont {J.-W.}\ \bibnamefont {Luo}}, \bibinfo {author} {\bibfnamefont {E.~I.}\ \bibnamefont {Rashba}},\ and\ \bibinfo {author} {\bibfnamefont {A.}~\bibnamefont {Zunger}},\ }\bibfield  {title} {\bibinfo {title} {Giant momentum-dependent spin splitting in centrosymmetric low-z antiferromagnets},\ }\href@noop {} {\bibfield  {journal} {\bibinfo  {journal} {Physical Review B}\ }\textbf {\bibinfo {volume} {102}},\ \bibinfo {pages} {014422} (\bibinfo {year} {2020})}\BibitemShut {NoStop}%
\bibitem [{\citenamefont {{\v{S}}mejkal}\ \emph {et~al.}(2020)\citenamefont {{\v{S}}mejkal}, \citenamefont {Gonz{\'a}lez-Hern{\'a}ndez}, \citenamefont {Jungwirth},\ and\ \citenamefont {Sinova}}]{vsmejkal2020crystal}%
  \BibitemOpen
  \bibfield  {author} {\bibinfo {author} {\bibfnamefont {L.}~\bibnamefont {{\v{S}}mejkal}}, \bibinfo {author} {\bibfnamefont {R.}~\bibnamefont {Gonz{\'a}lez-Hern{\'a}ndez}}, \bibinfo {author} {\bibfnamefont {T.}~\bibnamefont {Jungwirth}},\ and\ \bibinfo {author} {\bibfnamefont {J.}~\bibnamefont {Sinova}},\ }\bibfield  {title} {\bibinfo {title} {Crystal time-reversal symmetry breaking and spontaneous hall effect in collinear antiferromagnets},\ }\href@noop {} {\bibfield  {journal} {\bibinfo  {journal} {Science advances}\ }\textbf {\bibinfo {volume} {6}},\ \bibinfo {pages} {eaaz8809} (\bibinfo {year} {2020})}\BibitemShut {NoStop}%
\bibitem [{\citenamefont {Mazin}\ \emph {et~al.}(2021)\citenamefont {Mazin}, \citenamefont {Koepernik}, \citenamefont {Johannes}, \citenamefont {Gonz{\'a}lez-Hern{\'a}ndez},\ and\ \citenamefont {{\v{S}}mejkal}}]{mazin2021prediction}%
  \BibitemOpen
  \bibfield  {author} {\bibinfo {author} {\bibfnamefont {I.~I.}\ \bibnamefont {Mazin}}, \bibinfo {author} {\bibfnamefont {K.}~\bibnamefont {Koepernik}}, \bibinfo {author} {\bibfnamefont {M.~D.}\ \bibnamefont {Johannes}}, \bibinfo {author} {\bibfnamefont {R.}~\bibnamefont {Gonz{\'a}lez-Hern{\'a}ndez}},\ and\ \bibinfo {author} {\bibfnamefont {L.}~\bibnamefont {{\v{S}}mejkal}},\ }\bibfield  {title} {\bibinfo {title} {Prediction of unconventional magnetism in doped fesb2},\ }\href@noop {} {\bibfield  {journal} {\bibinfo  {journal} {Proceedings of the National Academy of Sciences}\ }\textbf {\bibinfo {volume} {118}},\ \bibinfo {pages} {e2108924118} (\bibinfo {year} {2021})}\BibitemShut {NoStop}%
\bibitem [{\citenamefont {Ryden}\ and\ \citenamefont {Lawson}(1970)}]{ryden1970magnetic}%
  \BibitemOpen
  \bibfield  {author} {\bibinfo {author} {\bibfnamefont {W.}~\bibnamefont {Ryden}}\ and\ \bibinfo {author} {\bibfnamefont {A.}~\bibnamefont {Lawson}},\ }\bibfield  {title} {\bibinfo {title} {Magnetic susceptibility of iro2 and ruo2},\ }\href@noop {} {\bibfield  {journal} {\bibinfo  {journal} {The Journal of Chemical Physics}\ }\textbf {\bibinfo {volume} {52}},\ \bibinfo {pages} {6058} (\bibinfo {year} {1970})}\BibitemShut {NoStop}%
\bibitem [{\citenamefont {Berlijn}\ \emph {et~al.}(2017)\citenamefont {Berlijn}, \citenamefont {Snijders}, \citenamefont {Delaire}, \citenamefont {Zhou}, \citenamefont {Maier}, \citenamefont {Cao}, \citenamefont {Chi}, \citenamefont {Matsuda}, \citenamefont {Wang}, \citenamefont {Koehler} \emph {et~al.}}]{berlijn2017itinerant}%
  \BibitemOpen
  \bibfield  {author} {\bibinfo {author} {\bibfnamefont {T.}~\bibnamefont {Berlijn}}, \bibinfo {author} {\bibfnamefont {P.~C.}\ \bibnamefont {Snijders}}, \bibinfo {author} {\bibfnamefont {O.}~\bibnamefont {Delaire}}, \bibinfo {author} {\bibfnamefont {H.-D.}\ \bibnamefont {Zhou}}, \bibinfo {author} {\bibfnamefont {T.~A.}\ \bibnamefont {Maier}}, \bibinfo {author} {\bibfnamefont {H.-B.}\ \bibnamefont {Cao}}, \bibinfo {author} {\bibfnamefont {S.-X.}\ \bibnamefont {Chi}}, \bibinfo {author} {\bibfnamefont {M.}~\bibnamefont {Matsuda}}, \bibinfo {author} {\bibfnamefont {Y.}~\bibnamefont {Wang}}, \bibinfo {author} {\bibfnamefont {M.~R.}\ \bibnamefont {Koehler}}, \emph {et~al.},\ }\bibfield  {title} {\bibinfo {title} {Itinerant antiferromagnetism in ruo 2},\ }\href@noop {} {\bibfield  {journal} {\bibinfo  {journal} {Physical review letters}\ }\textbf {\bibinfo {volume} {118}},\ \bibinfo {pages} {077201} (\bibinfo {year} {2017})}\BibitemShut {NoStop}%
\bibitem [{\citenamefont {Bai}\ \emph {et~al.}(2022)\citenamefont {Bai}, \citenamefont {Han}, \citenamefont {Feng}, \citenamefont {Zhou}, \citenamefont {Su}, \citenamefont {Wang}, \citenamefont {Liao}, \citenamefont {Zhu}, \citenamefont {Chen}, \citenamefont {Pan} \emph {et~al.}}]{bai2022observation}%
  \BibitemOpen
  \bibfield  {author} {\bibinfo {author} {\bibfnamefont {H.}~\bibnamefont {Bai}}, \bibinfo {author} {\bibfnamefont {L.}~\bibnamefont {Han}}, \bibinfo {author} {\bibfnamefont {X.}~\bibnamefont {Feng}}, \bibinfo {author} {\bibfnamefont {Y.}~\bibnamefont {Zhou}}, \bibinfo {author} {\bibfnamefont {R.}~\bibnamefont {Su}}, \bibinfo {author} {\bibfnamefont {Q.}~\bibnamefont {Wang}}, \bibinfo {author} {\bibfnamefont {L.}~\bibnamefont {Liao}}, \bibinfo {author} {\bibfnamefont {W.}~\bibnamefont {Zhu}}, \bibinfo {author} {\bibfnamefont {X.}~\bibnamefont {Chen}}, \bibinfo {author} {\bibfnamefont {F.}~\bibnamefont {Pan}}, \emph {et~al.},\ }\bibfield  {title} {\bibinfo {title} {Observation of spin splitting torque in a collinear antiferromagnet ruo 2},\ }\href@noop {} {\bibfield  {journal} {\bibinfo  {journal} {Physical Review Letters}\ }\textbf {\bibinfo {volume} {128}},\ \bibinfo {pages} {197202} (\bibinfo {year} {2022})}\BibitemShut {NoStop}%
\bibitem [{\citenamefont {Bose}\ \emph {et~al.}(2022)\citenamefont {Bose}, \citenamefont {Schreiber}, \citenamefont {Jain}, \citenamefont {Shao}, \citenamefont {Nair}, \citenamefont {Sun}, \citenamefont {Zhang}, \citenamefont {Muller}, \citenamefont {Tsymbal}, \citenamefont {Schlom} \emph {et~al.}}]{bose2022tilted}%
  \BibitemOpen
  \bibfield  {author} {\bibinfo {author} {\bibfnamefont {A.}~\bibnamefont {Bose}}, \bibinfo {author} {\bibfnamefont {N.~J.}\ \bibnamefont {Schreiber}}, \bibinfo {author} {\bibfnamefont {R.}~\bibnamefont {Jain}}, \bibinfo {author} {\bibfnamefont {D.-F.}\ \bibnamefont {Shao}}, \bibinfo {author} {\bibfnamefont {H.~P.}\ \bibnamefont {Nair}}, \bibinfo {author} {\bibfnamefont {J.}~\bibnamefont {Sun}}, \bibinfo {author} {\bibfnamefont {X.~S.}\ \bibnamefont {Zhang}}, \bibinfo {author} {\bibfnamefont {D.~A.}\ \bibnamefont {Muller}}, \bibinfo {author} {\bibfnamefont {E.~Y.}\ \bibnamefont {Tsymbal}}, \bibinfo {author} {\bibfnamefont {D.~G.}\ \bibnamefont {Schlom}}, \emph {et~al.},\ }\bibfield  {title} {\bibinfo {title} {Tilted spin current generated by the collinear antiferromagnet ruthenium dioxide},\ }\href@noop {} {\bibfield  {journal} {\bibinfo  {journal} {Nature Electronics}\ }\textbf {\bibinfo {volume} {5}},\ \bibinfo {pages} {267} (\bibinfo {year} {2022})}\BibitemShut {NoStop}%
\bibitem [{\citenamefont {Feng}\ \emph {et~al.}(2022)\citenamefont {Feng}, \citenamefont {Zhou}, \citenamefont {{\v{S}}mejkal}, \citenamefont {Wu}, \citenamefont {Zhu}, \citenamefont {Guo}, \citenamefont {Gonz{\'a}lez-Hern{\'a}ndez}, \citenamefont {Wang}, \citenamefont {Yan}, \citenamefont {Qin} \emph {et~al.}}]{feng2022anomalous}%
  \BibitemOpen
  \bibfield  {author} {\bibinfo {author} {\bibfnamefont {Z.}~\bibnamefont {Feng}}, \bibinfo {author} {\bibfnamefont {X.}~\bibnamefont {Zhou}}, \bibinfo {author} {\bibfnamefont {L.}~\bibnamefont {{\v{S}}mejkal}}, \bibinfo {author} {\bibfnamefont {L.}~\bibnamefont {Wu}}, \bibinfo {author} {\bibfnamefont {Z.}~\bibnamefont {Zhu}}, \bibinfo {author} {\bibfnamefont {H.}~\bibnamefont {Guo}}, \bibinfo {author} {\bibfnamefont {R.}~\bibnamefont {Gonz{\'a}lez-Hern{\'a}ndez}}, \bibinfo {author} {\bibfnamefont {X.}~\bibnamefont {Wang}}, \bibinfo {author} {\bibfnamefont {H.}~\bibnamefont {Yan}}, \bibinfo {author} {\bibfnamefont {P.}~\bibnamefont {Qin}}, \emph {et~al.},\ }\bibfield  {title} {\bibinfo {title} {An anomalous hall effect in altermagnetic ruthenium dioxide},\ }\href@noop {} {\bibfield  {journal} {\bibinfo  {journal} {Nature Electronics}\ }\textbf {\bibinfo {volume} {5}},\ \bibinfo {pages} {735} (\bibinfo {year} {2022})}\BibitemShut {NoStop}%
\bibitem [{\citenamefont {Guo}\ \emph {et~al.}(2024)\citenamefont {Guo}, \citenamefont {Zhang}, \citenamefont {Zhu}, \citenamefont {Jiang}, \citenamefont {Jiang}, \citenamefont {Wu}, \citenamefont {Dong}, \citenamefont {Xu}, \citenamefont {He}, \citenamefont {He} \emph {et~al.}}]{guo2024direct}%
  \BibitemOpen
  \bibfield  {author} {\bibinfo {author} {\bibfnamefont {Y.}~\bibnamefont {Guo}}, \bibinfo {author} {\bibfnamefont {J.}~\bibnamefont {Zhang}}, \bibinfo {author} {\bibfnamefont {Z.}~\bibnamefont {Zhu}}, \bibinfo {author} {\bibfnamefont {Y.-y.}\ \bibnamefont {Jiang}}, \bibinfo {author} {\bibfnamefont {L.}~\bibnamefont {Jiang}}, \bibinfo {author} {\bibfnamefont {C.}~\bibnamefont {Wu}}, \bibinfo {author} {\bibfnamefont {J.}~\bibnamefont {Dong}}, \bibinfo {author} {\bibfnamefont {X.}~\bibnamefont {Xu}}, \bibinfo {author} {\bibfnamefont {W.}~\bibnamefont {He}}, \bibinfo {author} {\bibfnamefont {B.}~\bibnamefont {He}}, \emph {et~al.},\ }\bibfield  {title} {\bibinfo {title} {Direct and inverse spin splitting effects in altermagnetic ruo2},\ }\href@noop {} {\bibfield  {journal} {\bibinfo  {journal} {Advanced Science}\ ,\ \bibinfo {pages} {2400967}} (\bibinfo {year} {2024})}\BibitemShut {NoStop}%
\bibitem [{\citenamefont {Karube}\ \emph {et~al.}(2022)\citenamefont {Karube}, \citenamefont {Tanaka}, \citenamefont {Sugawara}, \citenamefont {Kadoguchi}, \citenamefont {Kohda},\ and\ \citenamefont {Nitta}}]{karube2022observation}%
  \BibitemOpen
  \bibfield  {author} {\bibinfo {author} {\bibfnamefont {S.}~\bibnamefont {Karube}}, \bibinfo {author} {\bibfnamefont {T.}~\bibnamefont {Tanaka}}, \bibinfo {author} {\bibfnamefont {D.}~\bibnamefont {Sugawara}}, \bibinfo {author} {\bibfnamefont {N.}~\bibnamefont {Kadoguchi}}, \bibinfo {author} {\bibfnamefont {M.}~\bibnamefont {Kohda}},\ and\ \bibinfo {author} {\bibfnamefont {J.}~\bibnamefont {Nitta}},\ }\bibfield  {title} {\bibinfo {title} {Observation of spin-splitter torque in collinear antiferromagnetic ruo 2},\ }\href@noop {} {\bibfield  {journal} {\bibinfo  {journal} {Physical review letters}\ }\textbf {\bibinfo {volume} {129}},\ \bibinfo {pages} {137201} (\bibinfo {year} {2022})}\BibitemShut {NoStop}%
\bibitem [{\citenamefont {Liao}\ \emph {et~al.}(2024)\citenamefont {Liao}, \citenamefont {Wang}, \citenamefont {Tien}, \citenamefont {Huang},\ and\ \citenamefont {Qu}}]{liao2024separation}%
  \BibitemOpen
  \bibfield  {author} {\bibinfo {author} {\bibfnamefont {C.-T.}\ \bibnamefont {Liao}}, \bibinfo {author} {\bibfnamefont {Y.-C.}\ \bibnamefont {Wang}}, \bibinfo {author} {\bibfnamefont {Y.-C.}\ \bibnamefont {Tien}}, \bibinfo {author} {\bibfnamefont {S.-Y.}\ \bibnamefont {Huang}},\ and\ \bibinfo {author} {\bibfnamefont {D.}~\bibnamefont {Qu}},\ }\bibfield  {title} {\bibinfo {title} {Separation of inverse altermagnetic spin-splitting effect from inverse spin hall effect in ruo 2},\ }\href@noop {} {\bibfield  {journal} {\bibinfo  {journal} {Physical Review Letters}\ }\textbf {\bibinfo {volume} {133}},\ \bibinfo {pages} {056701} (\bibinfo {year} {2024})}\BibitemShut {NoStop}%
\bibitem [{\citenamefont {Liu}\ \emph {et~al.}(2023)\citenamefont {Liu}, \citenamefont {Bai}, \citenamefont {Song}, \citenamefont {Ji}, \citenamefont {Lou}, \citenamefont {Zhang}, \citenamefont {Song},\ and\ \citenamefont {Jin}}]{liu2023inverse}%
  \BibitemOpen
  \bibfield  {author} {\bibinfo {author} {\bibfnamefont {Y.}~\bibnamefont {Liu}}, \bibinfo {author} {\bibfnamefont {H.}~\bibnamefont {Bai}}, \bibinfo {author} {\bibfnamefont {Y.}~\bibnamefont {Song}}, \bibinfo {author} {\bibfnamefont {Z.}~\bibnamefont {Ji}}, \bibinfo {author} {\bibfnamefont {S.}~\bibnamefont {Lou}}, \bibinfo {author} {\bibfnamefont {Z.}~\bibnamefont {Zhang}}, \bibinfo {author} {\bibfnamefont {C.}~\bibnamefont {Song}},\ and\ \bibinfo {author} {\bibfnamefont {Q.}~\bibnamefont {Jin}},\ }\bibfield  {title} {\bibinfo {title} {Inverse altermagnetic spin splitting effect-induced terahertz emission in ruo2},\ }\href@noop {} {\bibfield  {journal} {\bibinfo  {journal} {Advanced Optical Materials}\ }\textbf {\bibinfo {volume} {11}},\ \bibinfo {pages} {2300177} (\bibinfo {year} {2023})}\BibitemShut {NoStop}%
\bibitem [{\citenamefont {Bai}\ \emph {et~al.}(2023)\citenamefont {Bai}, \citenamefont {Zhang}, \citenamefont {Zhou}, \citenamefont {Chen}, \citenamefont {Wan}, \citenamefont {Han}, \citenamefont {Zhu}, \citenamefont {Liang}, \citenamefont {Su}, \citenamefont {Han} \emph {et~al.}}]{bai2023efficient}%
  \BibitemOpen
  \bibfield  {author} {\bibinfo {author} {\bibfnamefont {H.}~\bibnamefont {Bai}}, \bibinfo {author} {\bibfnamefont {Y.}~\bibnamefont {Zhang}}, \bibinfo {author} {\bibfnamefont {Y.}~\bibnamefont {Zhou}}, \bibinfo {author} {\bibfnamefont {P.}~\bibnamefont {Chen}}, \bibinfo {author} {\bibfnamefont {C.}~\bibnamefont {Wan}}, \bibinfo {author} {\bibfnamefont {L.}~\bibnamefont {Han}}, \bibinfo {author} {\bibfnamefont {W.}~\bibnamefont {Zhu}}, \bibinfo {author} {\bibfnamefont {S.}~\bibnamefont {Liang}}, \bibinfo {author} {\bibfnamefont {Y.}~\bibnamefont {Su}}, \bibinfo {author} {\bibfnamefont {X.}~\bibnamefont {Han}}, \emph {et~al.},\ }\bibfield  {title} {\bibinfo {title} {Efficient spin-to-charge conversion via altermagnetic spin splitting effect in antiferromagnet ruo 2},\ }\href@noop {} {\bibfield  {journal} {\bibinfo  {journal} {Physical review letters}\ }\textbf {\bibinfo {volume} {130}},\ \bibinfo {pages} {216701} (\bibinfo {year} {2023})}\BibitemShut {NoStop}%
\bibitem [{\citenamefont {Fedchenko}\ \emph {et~al.}(2024)\citenamefont {Fedchenko}, \citenamefont {Min{\'a}r}, \citenamefont {Akashdeep}, \citenamefont {D’Souza}, \citenamefont {Vasilyev}, \citenamefont {Tkach}, \citenamefont {Odenbreit}, \citenamefont {Nguyen}, \citenamefont {Kutnyakhov}, \citenamefont {Wind} \emph {et~al.}}]{fedchenko2024observation}%
  \BibitemOpen
  \bibfield  {author} {\bibinfo {author} {\bibfnamefont {O.}~\bibnamefont {Fedchenko}}, \bibinfo {author} {\bibfnamefont {J.}~\bibnamefont {Min{\'a}r}}, \bibinfo {author} {\bibfnamefont {A.}~\bibnamefont {Akashdeep}}, \bibinfo {author} {\bibfnamefont {S.~W.}\ \bibnamefont {D’Souza}}, \bibinfo {author} {\bibfnamefont {D.}~\bibnamefont {Vasilyev}}, \bibinfo {author} {\bibfnamefont {O.}~\bibnamefont {Tkach}}, \bibinfo {author} {\bibfnamefont {L.}~\bibnamefont {Odenbreit}}, \bibinfo {author} {\bibfnamefont {Q.}~\bibnamefont {Nguyen}}, \bibinfo {author} {\bibfnamefont {D.}~\bibnamefont {Kutnyakhov}}, \bibinfo {author} {\bibfnamefont {N.}~\bibnamefont {Wind}}, \emph {et~al.},\ }\bibfield  {title} {\bibinfo {title} {Observation of time-reversal symmetry breaking in the band structure of altermagnetic ruo2},\ }\href@noop {} {\bibfield  {journal} {\bibinfo  {journal} {Science advances}\ }\textbf {\bibinfo {volume} {10}},\ \bibinfo {pages} {eadj4883} (\bibinfo {year} {2024})}\BibitemShut {NoStop}%
\bibitem [{\citenamefont {Hiraishi}\ \emph {et~al.}(2024)\citenamefont {Hiraishi}, \citenamefont {Okabe}, \citenamefont {Koda}, \citenamefont {Kadono}, \citenamefont {Muroi}, \citenamefont {Hirai},\ and\ \citenamefont {Hiroi}}]{hiraishi2024nonmagnetic}%
  \BibitemOpen
  \bibfield  {author} {\bibinfo {author} {\bibfnamefont {M.}~\bibnamefont {Hiraishi}}, \bibinfo {author} {\bibfnamefont {H.}~\bibnamefont {Okabe}}, \bibinfo {author} {\bibfnamefont {A.}~\bibnamefont {Koda}}, \bibinfo {author} {\bibfnamefont {R.}~\bibnamefont {Kadono}}, \bibinfo {author} {\bibfnamefont {T.}~\bibnamefont {Muroi}}, \bibinfo {author} {\bibfnamefont {D.}~\bibnamefont {Hirai}},\ and\ \bibinfo {author} {\bibfnamefont {Z.}~\bibnamefont {Hiroi}},\ }\bibfield  {title} {\bibinfo {title} {Nonmagnetic ground state in ruo 2 revealed by muon spin rotation},\ }\href@noop {} {\bibfield  {journal} {\bibinfo  {journal} {Physical Review Letters}\ }\textbf {\bibinfo {volume} {132}},\ \bibinfo {pages} {166702} (\bibinfo {year} {2024})}\BibitemShut {NoStop}%
\bibitem [{\citenamefont {Ke{\ss}ler}\ \emph {et~al.}(2024)\citenamefont {Ke{\ss}ler}, \citenamefont {Garcia-Gassull}, \citenamefont {Suter}, \citenamefont {Prokscha}, \citenamefont {Salman}, \citenamefont {Khalyavin}, \citenamefont {Manuel}, \citenamefont {Orlandi}, \citenamefont {Mazin}, \citenamefont {Valent{\'\i}} \emph {et~al.}}]{kessler2024absence}%
  \BibitemOpen
  \bibfield  {author} {\bibinfo {author} {\bibfnamefont {P.}~\bibnamefont {Ke{\ss}ler}}, \bibinfo {author} {\bibfnamefont {L.}~\bibnamefont {Garcia-Gassull}}, \bibinfo {author} {\bibfnamefont {A.}~\bibnamefont {Suter}}, \bibinfo {author} {\bibfnamefont {T.}~\bibnamefont {Prokscha}}, \bibinfo {author} {\bibfnamefont {Z.}~\bibnamefont {Salman}}, \bibinfo {author} {\bibfnamefont {D.}~\bibnamefont {Khalyavin}}, \bibinfo {author} {\bibfnamefont {P.}~\bibnamefont {Manuel}}, \bibinfo {author} {\bibfnamefont {F.}~\bibnamefont {Orlandi}}, \bibinfo {author} {\bibfnamefont {I.~I.}\ \bibnamefont {Mazin}}, \bibinfo {author} {\bibfnamefont {R.}~\bibnamefont {Valent{\'\i}}}, \emph {et~al.},\ }\bibfield  {title} {\bibinfo {title} {Absence of magnetic order in ruo2: insights from $\mu$ sr spectroscopy and neutron diffraction},\ }\href@noop {} {\bibfield  {journal} {\bibinfo  {journal} {npj Spintronics}\ }\textbf {\bibinfo {volume} {2}},\ \bibinfo {pages} {50} (\bibinfo {year} {2024})}\BibitemShut {NoStop}%
\bibitem [{\citenamefont {Liu}\ \emph {et~al.}(2024)\citenamefont {Liu}, \citenamefont {Zhan}, \citenamefont {Li}, \citenamefont {Liu}, \citenamefont {Cheng}, \citenamefont {Shi}, \citenamefont {Deng}, \citenamefont {Zhang}, \citenamefont {Li}, \citenamefont {Ding} \emph {et~al.}}]{liu2024absence}%
  \BibitemOpen
  \bibfield  {author} {\bibinfo {author} {\bibfnamefont {J.}~\bibnamefont {Liu}}, \bibinfo {author} {\bibfnamefont {J.}~\bibnamefont {Zhan}}, \bibinfo {author} {\bibfnamefont {T.}~\bibnamefont {Li}}, \bibinfo {author} {\bibfnamefont {J.}~\bibnamefont {Liu}}, \bibinfo {author} {\bibfnamefont {S.}~\bibnamefont {Cheng}}, \bibinfo {author} {\bibfnamefont {Y.}~\bibnamefont {Shi}}, \bibinfo {author} {\bibfnamefont {L.}~\bibnamefont {Deng}}, \bibinfo {author} {\bibfnamefont {M.}~\bibnamefont {Zhang}}, \bibinfo {author} {\bibfnamefont {C.}~\bibnamefont {Li}}, \bibinfo {author} {\bibfnamefont {J.}~\bibnamefont {Ding}}, \emph {et~al.},\ }\bibfield  {title} {\bibinfo {title} {Absence of altermagnetic spin splitting character in rutile oxide ruo 2},\ }\href@noop {} {\bibfield  {journal} {\bibinfo  {journal} {Physical Review Letters}\ }\textbf {\bibinfo {volume} {133}},\ \bibinfo {pages} {176401} (\bibinfo {year} {2024})}\BibitemShut {NoStop}%
\bibitem [{\citenamefont {Wu}\ \emph {et~al.}(2021)\citenamefont {Wu}, \citenamefont {Yaw~Ameyaw}, \citenamefont {Doty},\ and\ \citenamefont {Jungfleisch}}]{wu2021principles}%
  \BibitemOpen
  \bibfield  {author} {\bibinfo {author} {\bibfnamefont {W.}~\bibnamefont {Wu}}, \bibinfo {author} {\bibfnamefont {C.}~\bibnamefont {Yaw~Ameyaw}}, \bibinfo {author} {\bibfnamefont {M.~F.}\ \bibnamefont {Doty}},\ and\ \bibinfo {author} {\bibfnamefont {M.~B.}\ \bibnamefont {Jungfleisch}},\ }\bibfield  {title} {\bibinfo {title} {Principles of spintronic thz emitters},\ }\href@noop {} {\bibfield  {journal} {\bibinfo  {journal} {Journal of Applied Physics}\ }\textbf {\bibinfo {volume} {130}} (\bibinfo {year} {2021})}\BibitemShut {NoStop}%
\bibitem [{\citenamefont {Zhang}\ \emph {et~al.}(2023)\citenamefont {Zhang}, \citenamefont {Cui}, \citenamefont {Wang}, \citenamefont {Chen}, \citenamefont {Liu}, \citenamefont {Qin}, \citenamefont {Guan}, \citenamefont {Tian}, \citenamefont {Yuan}, \citenamefont {Zhou} \emph {et~al.}}]{zhang2023nonrelativistic}%
  \BibitemOpen
  \bibfield  {author} {\bibinfo {author} {\bibfnamefont {S.}~\bibnamefont {Zhang}}, \bibinfo {author} {\bibfnamefont {Y.}~\bibnamefont {Cui}}, \bibinfo {author} {\bibfnamefont {S.}~\bibnamefont {Wang}}, \bibinfo {author} {\bibfnamefont {H.}~\bibnamefont {Chen}}, \bibinfo {author} {\bibfnamefont {Y.}~\bibnamefont {Liu}}, \bibinfo {author} {\bibfnamefont {W.}~\bibnamefont {Qin}}, \bibinfo {author} {\bibfnamefont {T.}~\bibnamefont {Guan}}, \bibinfo {author} {\bibfnamefont {C.}~\bibnamefont {Tian}}, \bibinfo {author} {\bibfnamefont {Z.}~\bibnamefont {Yuan}}, \bibinfo {author} {\bibfnamefont {L.}~\bibnamefont {Zhou}}, \emph {et~al.},\ }\bibfield  {title} {\bibinfo {title} {Nonrelativistic and nonmagnetic terahertz-wave generation via ultrafast current control in anisotropic conductive heterostructures},\ }\href@noop {} {\bibfield  {journal} {\bibinfo  {journal} {Advanced Photonics}\ }\textbf {\bibinfo {volume} {5}},\ \bibinfo {pages} {056006} (\bibinfo {year} {2023})}\BibitemShut {NoStop}%
\bibitem [{\citenamefont {Wu}\ \emph {et~al.}(2024)\citenamefont {Wu}, \citenamefont {Acuna}, \citenamefont {Huang}, \citenamefont {Wang}, \citenamefont {Gundlach}, \citenamefont {Doty}, \citenamefont {Zide},\ and\ \citenamefont {Jungfleisch}}]{wu2024hybrid}%
  \BibitemOpen
  \bibfield  {author} {\bibinfo {author} {\bibfnamefont {W.}~\bibnamefont {Wu}}, \bibinfo {author} {\bibfnamefont {W.}~\bibnamefont {Acuna}}, \bibinfo {author} {\bibfnamefont {Z.}~\bibnamefont {Huang}}, \bibinfo {author} {\bibfnamefont {X.}~\bibnamefont {Wang}}, \bibinfo {author} {\bibfnamefont {L.}~\bibnamefont {Gundlach}}, \bibinfo {author} {\bibfnamefont {M.~F.}\ \bibnamefont {Doty}}, \bibinfo {author} {\bibfnamefont {J.~M.}\ \bibnamefont {Zide}},\ and\ \bibinfo {author} {\bibfnamefont {M.~B.}\ \bibnamefont {Jungfleisch}},\ }\bibfield  {title} {\bibinfo {title} {Hybrid terahertz emitter for pulse shaping and chirality control},\ }\href@noop {} {\bibfield  {journal} {\bibinfo  {journal} {arXiv preprint arXiv:2406.05875}\ } (\bibinfo {year} {2024})}\BibitemShut {NoStop}%
\bibitem [{\citenamefont {Sato}\ \emph {et~al.}(2013)\citenamefont {Sato}, \citenamefont {Higuchi}, \citenamefont {Kanda}, \citenamefont {Konishi}, \citenamefont {Yoshioka}, \citenamefont {Suzuki}, \citenamefont {Misawa},\ and\ \citenamefont {Kuwata-Gonokami}}]{sato2013terahertz}%
  \BibitemOpen
  \bibfield  {author} {\bibinfo {author} {\bibfnamefont {M.}~\bibnamefont {Sato}}, \bibinfo {author} {\bibfnamefont {T.}~\bibnamefont {Higuchi}}, \bibinfo {author} {\bibfnamefont {N.}~\bibnamefont {Kanda}}, \bibinfo {author} {\bibfnamefont {K.}~\bibnamefont {Konishi}}, \bibinfo {author} {\bibfnamefont {K.}~\bibnamefont {Yoshioka}}, \bibinfo {author} {\bibfnamefont {T.}~\bibnamefont {Suzuki}}, \bibinfo {author} {\bibfnamefont {K.}~\bibnamefont {Misawa}},\ and\ \bibinfo {author} {\bibfnamefont {M.}~\bibnamefont {Kuwata-Gonokami}},\ }\bibfield  {title} {\bibinfo {title} {Terahertz polarization pulse shaping with arbitrary field control},\ }\href@noop {} {\bibfield  {journal} {\bibinfo  {journal} {Nature Photonics}\ }\textbf {\bibinfo {volume} {7}},\ \bibinfo {pages} {724} (\bibinfo {year} {2013})}\BibitemShut {NoStop}%
\end{thebibliography}%
\end{document}